# On-Chip High Extinction Ratio Single-Stage Mach-Zehnder Interferometer based on Multimode Interferometer

Shengjie Xie[1], Sylvain Veilleux[2], Mario Dagenais[1,*]

1.Department of Electrical and Computer Engineering, University of Maryland, College Park, MD 20742

2. Department of Astronomy, University of Maryland, College Park, MD 20742

*dage@umd.edu

**Abstract:** On-chip high extinction ratio Mach-Zehnder interferometers (MZI) have always attracted interest from researchers as it can be used in many applications in astrophotonics, optical switching, programmable photonic circuits, and quantum information. However, in previous research studies, ultra-high extinction ratio on-chip MZIs have only been achieved by using a multi-stage MZI approach. In this paper, we investigate a high extinction ratio single-stage MZI based on two cascaded multimode interferometers (MMI). We determine that TM noise is an important factor that can prevent us from achieving a high extinction ratio MZI. By introducing a bend-based TM filter without additional loss, we experimentally demonstrate that such a TM filter can improve the maximum extinction ratio of the MMI-MZI by more than 10 dB. With the TM filter, we report a record high 61.2 dB extinction ratio in a single stage, thermally tunable MMI-MZI with only 1.5 dB insertion loss and more than 60nm bandwidth. These results pave the way for many interesting applications.

**Introduction**

On-chip high performance Mach-Zehnder interferometer (MZI) have recently attracted much interest as it is a key component for many photonics applications such as in astrophotonics [1,2], optical switching [3,4] and programmable photonic circuits [5,6], and quantum information [7,8]. Among the figure of merits of a successful MZI, the extinction ratio is the most important factor in many situations. In astrophotonics, astronomers have used nulling interferometers, which are based on MZIs, to detect exoplanets. Since the host star is at least six orders of magnitude brighter than the exoplanet, the light from the host star must be removed, either by using a coronagraph or an interferometer. In this case, a higher extinction ratio for the interferometer directly translates into better capability for finding exoplanets. In programmable photonic circuits, MZI is widely used to change the optical routing. A higher extinction ratio represents a better control precision of the photonic circuits. In quantum computation, MZI is often used for qubits manipulation. A MZI with higher extinction ratio can greatly increase the fidelity of quantum computation. In optical switching, a high extinction ratio MZI can also benefit optical routing and optical modulation and thus improve the overall performance of the photonic circuits. Conventionally, researchers found that due to fabrication imperfection, the beam splitter in a MZI, either a directional coupler or a multimode interferometer (MMI), will unevenly split the injected light. As a result, the extinction ratio is limited. A common solution is to use a multi-stage MZI. The first stage MZI works as a reconfigurable beam splitter to ensure a 50:50 splitting ratio so that the second stage MZI can achieve a high extinction ratio [8–10]. However, this method is not ideal since it requires the simultaneous control of two separate MZIs, which increases the control complexity and the power consumption due to the additional phase shifters.

In this paper, unlike the previously reported multi-stage high extinction ratio MZI, we report a single-stage high extinction ratio MMI-MZI on silicon nitride ($Si_3N_4$), a material that has been shown to be an ultra low loss material for photonics applications [11–14]. We first optimize the MMI design to maximize the extinction ratio of the MMI-MZI. A noise mechanism due to unwanted TM light is investigated. We design a bend-based TM filter, and we characterize the suppression raito of the TM filter with a Bragg grating. A more than 25 dB TM light suppression is achieved with the proposed TM filter. With the on-chip TM filter, we demonstrate a thermally tunable MMI-MZI with an insertion loss of 1.5dB, a maximum extinction ratio of 61.2 dB, and a tuning power of 200mW working in C+L band. Lastly, by comparing the extinction ratio of the MMI-MZI with or without TM filter, we experimentally show that the TM filter can improve the maximum extinction ratio of the MMI-MZI by more than 10 dB. All of these technical advancements pave the way for future applications in astrophotonics, optical switching, programmable photonic circuits, and quantum information.

**MMI-MZI Design**

Fig. 1(a) shows the schematic of the MZI. The proposed MZI consists of two identical 2×2 MMIs and two arms, both of which are thermally tunable to maximize the tuning range of the MZI. Compared to our previous work [15], the MMI is designed for 300nm thick $Si_3N_4$ for a balance of low loss and compactness. The cross section of the waveguide is shown in Fig. 1(b). The top oxide and buried oxide thickness are set to be $4\mu m$ and $5\mu m$ respectively to minimize the mode interaction with the silicon substrate and the metal heater. There are several considerations in designing a high extinction ratio MMI-MZI before we start the simulation. While simultaneously achieving broadband and high extinction ratio MZI are highly desirable characteristics for many applications, it is impossible in practice due to the nature of the MMI. The bandwidth of the MZI purely depends on the bandwidth of a single MMI. The bandwidth of a MMI is related to the MMI's dimension:

$$\frac{\delta \lambda_0}{\lambda_0} \approx \frac{\delta L}{L} = \frac{2\delta W_e}{W_e} \qquad (1)$$

where the $\lambda_0$ is the central wavelength of the MMI, $L$ is the length of the MMI, and $W_e$ is the effective width of the MMI, which can be written as [16]

$$W_e = W + \left(\frac{\lambda_0}{\pi}\right)(n_r^2 - n_c^2)^{-\frac{1}{2}} \qquad (2)$$

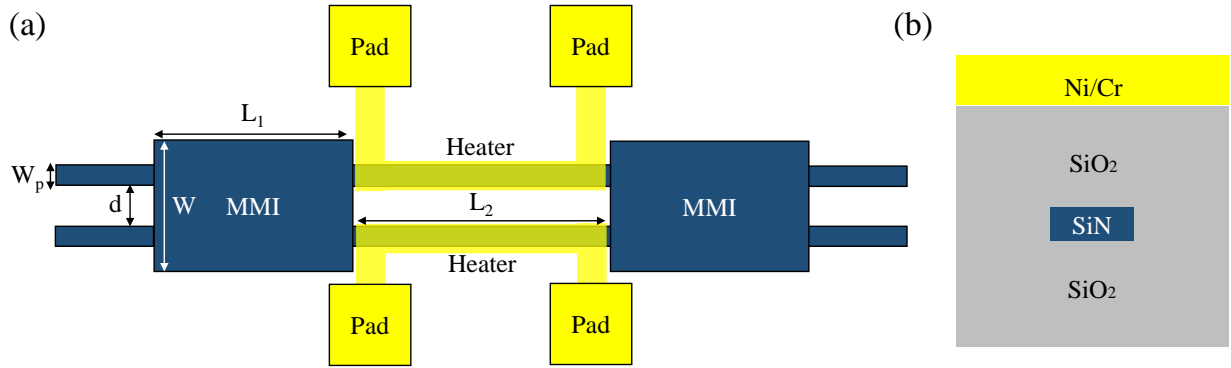

Figure 1. (a) Schematics of MZI. (b) Cross section of the $Si_3N_4$ waveguides.

From Equations 1 and 2, the bandwidth decreases as the dimension of the MMI is increased. However, MMI works based on the "self-imaging" principle [16]. The more modes the MMI supports, the better the imaging quality will be, thus resulting in a higher extinction ratio. Hence, there is a fundamental tradeoff between a larger bandwidth and a higher extinction ratio. Fig. 2 (a-b) shows the simulated transmission of the MMI. When the width of the MMI is set to be $10\mu m$, supporting 8 guided TE mode, the MMI has a 3-dB bandwidth larger than 200nm. But the largest extinction ratio is only ~40dB. Instead, a MMI with a width of $15\mu m$, supporting 10 guided modes, has an extinction ratio of better than 65 dB with a reduced 3-dB bandwidth of 80nm. Despite the fundamental tradeoff, we put our effort into improving the extinction ratio in this study.

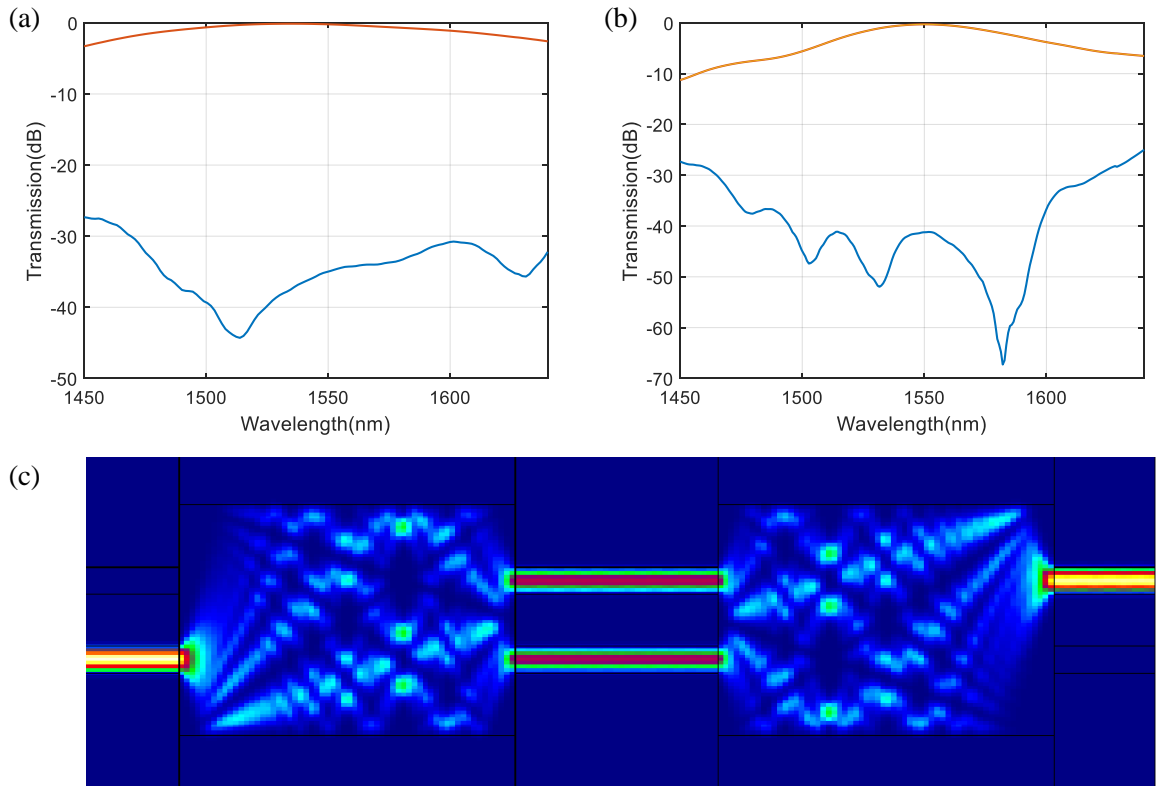

Figure 2 (a) Simulation results of the MZI with 10 $\mu m$ MMI transmission at the null port and anti-null port. (b) Simulation results of the MZI with 15 $\mu m$ MMI transmission at the null port and anti-null port. (c) Simulated propagation profile of the MZI.

After careful simulation, we design the dimension of the MMI to be 15×166.23 $\mu m^2$. The port width $W_p$ is designed to be 1.8 $\mu m$ with a separation $d$ of 3.2 $\mu m$ between the two ports of the MMI. The MMI is optimized for TE polarized light. Such design enables the paired interference mode [15,16], so that all modes whose mode number is an integer multiple of 3 are not excited. As a result, the MMI dimension is reduced three-fold. The FIMMPROP simulated transmission at two output ports and the mode profile are shown in Fig. 2 (c). Since we need $2\pi$ phase shift capability for the MZI and since the thermo-optic coefficient of both $Si_3N_4$ and $SiO_2$ are small (2.4×10[-5]/K and 0.96×10[-5]/K, respectively) [17], the two MMIs are separated by 600 $\mu m$, and the heater are designed to be 500×10 $\mu m^2$ to allow larger thermal tuning capability.

**TM noise reduction**

While other researchers mainly put their efforts into mitigating the impact of fabrication imperfection, such as multi-stage MZI, to improve the extinction ratio, other factor that limit the extinction ratio has long been ignored. Indeed, in addition to the power imbalance due to fabrication imperfections, we found that TM noise is an important factor that prevents us from getting a high extinction ratio MMI-MZI. Since the MMI is optimized for TE polarization, when TM polarized light goes into the MMI, a small portion of the TM light will still go to the port that should have minimal output, which will limit the overall extinction ratio. As shown in Fig. 3 (a-b), for the MMI that is optimized for TE mode, the light does not perfectly converge in the output port with a TM input, and the simulated extinction ratio decreases from 65 dB to around 40 dB. Unlike the fabrication imperfection of the MZI, which can be minimized by advanced fabrication techniques, such as high precision e-beam lithography and dry etch, the TM noise issue is rooted in the MMI design. Due to the limited polarization extinction ratio of the laser and polarization maintaining fiber, an on-chip broadband TM filter is the only solution to this issue. Here, we take advantage of the different bending loss sensitivity of the TE and TM modes to remove the TM noise [18]. As shown in Fig. 3 (c), we design a bent waveguide with a cross section of 900×300 nm, operating in single mode condition. As the bending radius decreases, the bending loss increases significantly for both the TE mode and TM mode. However, with a bending radius of 50 $\mu m$, while the bending loss for TM mode is 4.8 dB/360° bend, the bending loss for TE mode remains negligible. With the significant bending loss difference between TE and TM modes, we can implement on-chip broadband TM filtering. We test the performance of the TM filter with a Bragg grating. The Bragg grating is defined by a sidewall corrugation method with a width of 900nm and 1200nm, respectively. It is designed to have a stopband at around 1600 nm for TE input. As shown in Fig. 3 (d), without the broadband TM filter, the rejection ratio of the Bragg grating is limited by the TM noise. The TM noise floor is observed to be at around -20 dB. After adding the TM filter to the Bragg grating, the TM noise is suppressed by 25 dB with 720° bends. In fact, the TM suppression ratio is limited by the depth of the stopband and can be even larger. It is also worth noting that the transmission of the Bragg grating (with or without TM filter) outside of the stopband overlaps with each other, which is a good indicator that the bending loss of the TE mode is negligible. The stronger than simulated TM suppression is attributed to the fact that the bending loss simulation does not take the mode mismatch loss between the straight waveguide mode and the bend mode into consideration. Thus, the experimental TM suppression can be higher than the simulated results. In the following high extinction ratio MZI experiment, a TM filter with 720° 50 $\mu m$ radius bend is always included in all MZI designs to remove the on-chip TM noise.

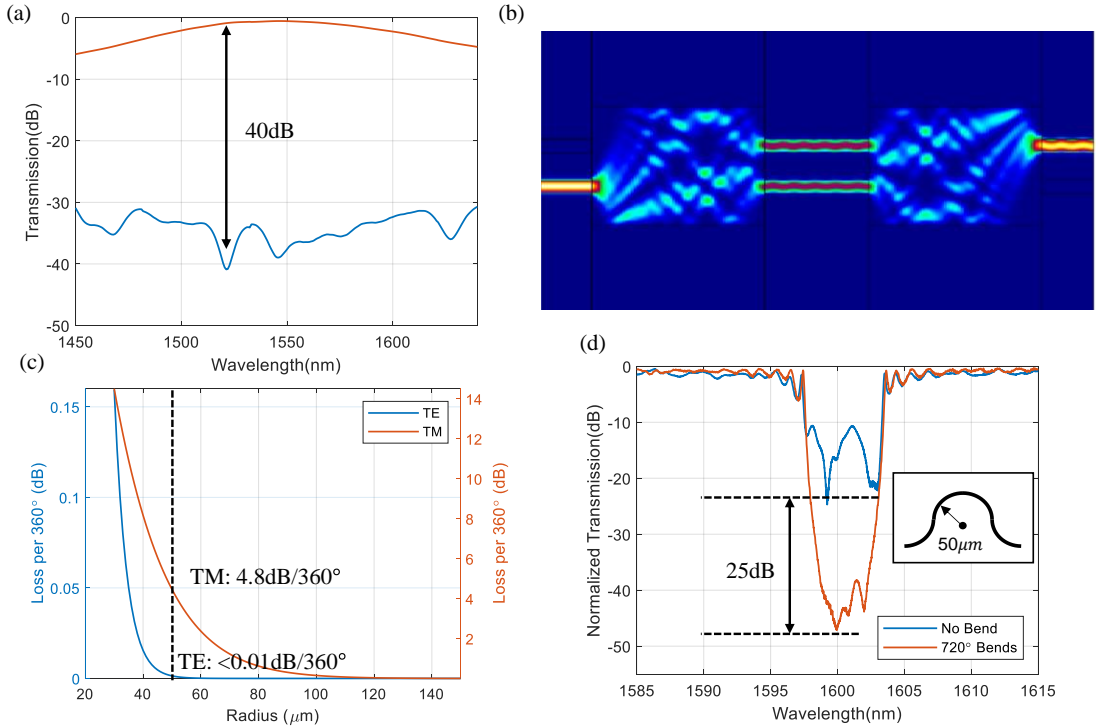

Figure 3. (a) Simulated transmission of the 15$\mu m$ wide MZI, which is optimized for TE input, with a TM input. (b) Simulated mode propagation profile of the 15$\mu m$ wide MZI with a TM input. (c) Simulated TE and TM mode bending losses. (d) Experimental results of the bend waveguide TM filter. Inset: schematic of the bending waveguides used in the test.

**Experimental Result**

As shown in Fig. 4(a), the fabrication of the MMI-MZI starts from a silicon wafer with a 5$\mu m$ thermally grown silicon dioxide layer. Then 300 nm thick $Si_3N_4$ is deposited using low-pressure chemical vapor deposition (LPCVD). A positive tone e-beam resist, ZEP-520A, is exposed using a Elionix ELS G-100 100kV e-beam lithography system. After chromium deposition and lift off, the e-beam pattern is transferred to the $Si_3N_4$ layer by inductively coupled plasma (ICP) etch. A 4$\mu m$ top oxide cladding is then deposited by plasma enhanced chemical vapor deposition (PECVD). We use some e-beam overlay feature to align the heater pattern to the arms between the two MMIs. Finally, 130nm Nichrome metal is deposited on top of the top oxide as the heater. Fig. 4(b) shows the SEM images of the fabricated devices. The device's dimension agrees well with the design. Thanks to the precision of the e-beam overlay process, the heater aligns well with the arms of the MZI, with a precision better than $\pm 1\mu m$, as shown in Fig. 4 (c), which ensures an efficient thermal tuning of the MZI.

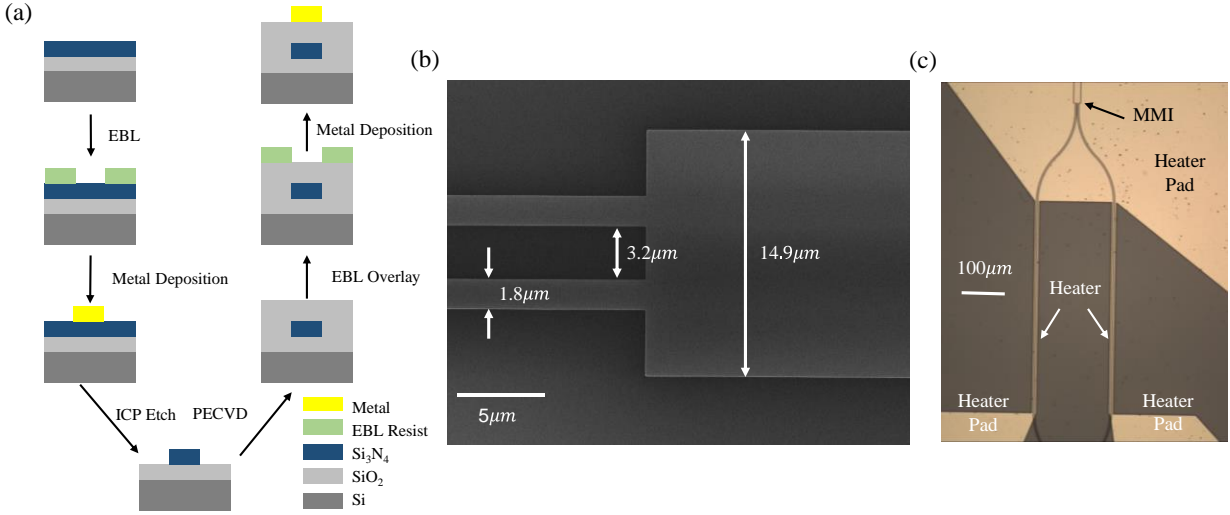

Figure 4. (a) Schematic of the fabrication process. (b) SEM image of the fabricated MMI. (c) Microscopic image of the fabricated MMI-MZI. The second MMI is not shown due to limited field of view of the microscope.

To characterize the performance of the MZI, a TE polarized tunable laser source with a tuning range of 1450-1640 nm and 5 pm scanning resolution is used. The light is edge coupled in and out of the chip by polarization maintaining fiber with a coupling efficiency of 4dB/facet. The signal is then analyzed by an optical powermeter. The heater is controlled by a Keithley 2401 sourcemeter with a current precision of better than $1\mu A$.

The measurement result of the MMI-MZI is shown in Fig. 5. Fig. 5 (a) shows the transmission of the MMI-MZI at different heater power. The transmission is normalized to a reference straight waveguide to remove the coupling loss and waveguide propagation loss. The insertion loss of the MMI-MZI is measured to be 1.5 dB at 1550 nm, and the 3-dB bandwidth of the MMI is measured to be ~80nm, and agrees well with simulation results as shown in Fig. 2(b). By applying power to the heater, the transmission of the MZI gradually shifts from the maximum state to the minimum state. As shown in Fig. 5 (b), with a tuning power of 200 mW, the minimum state is achieved, and a maximum extinction ratio of 61.2 dB is observed at 1562 nm, which agrees reasonably well with the simulation results. In Fig. 5(c), we demonstrate that by slightly tuning the heater power by less than 1 mW, the maximum extinction ratio can be achieved at different wavelengths.. As shown in Fig. 5(d), the 3-dB bandwidth of the high extinction ratio MMI-MZI is around 60nm, which is primarily limited by the 3-dB bandwidth of the MMI. We also experimentally investigate the performance of the MZI without an on-chip TM filter. Fig. 5 (e) shows the transmission of the MZI without an on-chip TM filter. Without the on-chip TM filter, with a TE input, the maximum extinction ratio of the MZI is limited to less than 50 dB, proving that the on-chip TM filter is critical to achieve a high extinction ratio MZI. We also demonstrate the switching behavior of the MZI by tuning the power around the maximum and minimum state at 1562nm, as shown in Fig. 5 (f)

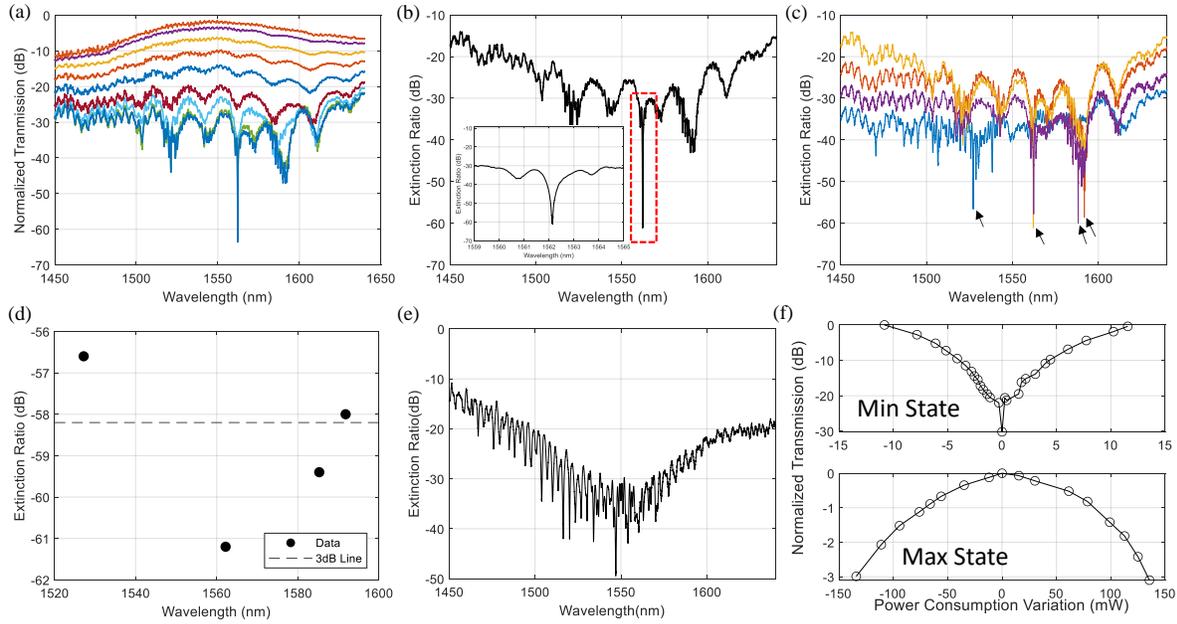

Figure 5. Experimental results of the MMI-MZI. (a) Normalized transmission of the MMI-MZI for different tuning powers from 12.35 mW to 216.55 mW. (b) Extinction ratio of the MMI-MZI. Inset: Zoomed-in view of the extinction ratio around 1560 nm. (c) MMI-MZI transmission with maximum ER achieved at different wavelengths by tuning the heater power for less than 1 mW. (d) Maximum extinction ratio achieved at different wavelength, corresponding to the arrows shown in (c). (e) Transmission of the MMI-MZI without TM filter. (f) Switching behavior of the MMI-MZI around the maximum state and minimum state.

Table 1. Comparison of several experimentally demonstrated MZI

| Structure | Material | Extinction Ratio | Insertion Loss | Bandwidth | Reference |
|---|---|---|---|---|---|
| 2-stage DC-MZI | LiNbO3 | 53dB | 3dB/cm | 100nm | [9] |
| 2-stage DC-MZI | SOI | >65dB | n/a | C-band 160nm O-band 95nm | [10] |
| 3-stage MMI-MZ | SOI | 60.5dB | n/a | n/a | [8] |
| MMI-MZI | SOI | >20dB | 1-4dB | 65nm | [19] |
| MMI-MZI | $Si_3N_4$ | 48dB | <5dB | 80nm | [20] |
| MMI-MZI | SOI | 30dB | 0.5dB | ~100nm | [21] |
| DC-MZI | SOI | 38dB | 1.0dB | n/a | [22] |
| DC-MZI | SOI | 35dB | n/a | n/a | [23] |
| MMI-MZI | $Si_3N_4$ | 46.1±2.5dB | n/a | ~100nm | [5] |
| MMI-MZI | $Si_3N_4$ | 61.2dB | 1.5dB | 60nm | This work |

**Discussion and Conclusion**

A comparison of the performance of several MZI using different structures and materials are shown in Table 1. Thanks to the bending waveguide based TM filter and ultra high precision of the e-beam lithography, the MZI demonstrated in this work shows a much higher extinction ratio than other previously demonstrated single-stage MZI, and a extinction ratio comparable to the results obtained from the best-to-date multi-stage MZIs. In addition, the MMI-MZI demonstrated in this work shows a reasonably low insertion loss due to the low-loss nature of properly designed $Si_3N_4$ photonic components. There are multiple directions to further improve the performance of the high extinction ratio MMI-MZI. Firstly, the tuning power can be greatly reduced in future work by optimizing the heater design and using a more efficient heater material [24]. Heat dissipation may be avoided altogether by using PZT materials for tuning the interferometer [25]. Secondly, the response time of the MZI can be reduced by removing the silicon substrate as reported in [21], or again, by using PZT tuning. Lastly, by designing a wider MMI, the extinction ratio can be improved further.

In summary, we have experimentally studied a single-stage, thermally tunable, high extinction ratio $Si_3N_4$ MMI-MZI. The MMI-MZI has an insertion loss of 1.5 dB at 1550 nm and a 3-dB bandwidth of ~60nm. Such a single-stage MMI-MZI has great potential for reducing the control complexity and power consumption. We found that on-chip TM noise is an additional noise mechanism. After removing the TM noise by using a carefully designed broadband TM filter, we demonstrate that a tuning power of 200mW combined with a record high extinction ratio of 61.2 dB can be achieved at 1562nm. We also experimentally investigate the extinction ratio of the MZI without a TM filter and determined that the on-chip TM filter can improve the maximum extinction ratio of the MMI-MZI by more than 10 dB, which proves the importance of removing TM noise for an MZI with a high extinction ratio. We envision that the high extinction ratio MMI-MZI will benefit many applications in astrophotonics, optical swtching, programmable photonic circuits, and quantum information.

**Acknowledgement**

The authors acknowledge the support from NASA (contract # 16-APRA16-0064). The authors thank Jiahao Zhan and Xiheng Ai of the University of Maryland for stimulating discussions. The authors would also like to acknowledge the engineering team at the Maryland Nanocenter for fabrication support.

**Reference**
1. M. A. Martinod, B. Norris, P. Tuthill, T. Lagadec, N. Jovanovic, N. Cvetojevic, S. Gross, A. Arriola, T. Gretzinger, M. J. Withford, O. Guyon, J. Lozi, S. Vievard, V. Deo, J. S. Lawrence, and S. Leon-Saval, "Scalable photonic-based nulling interferometry with the dispersed multi-baseline GLINT instrument," Nat. Commun. 121 **12**, 1–11 (2021).
2. B. R. M. Norris, N. Cvetojevic, T. Lagadec, N. Jovanovic, S. Gross, A. Arriola, T. Gretzinger, M. A. Martinod, O. Guyon, J. Lozi, M. J. Withford, J. S. Lawrence, and P. Tuthill, "First on-sky demonstration of an integrated-photonic nulling interferometer: the GLINT instrument," Mon. Not. R. Astron. Soc. **491**, 4180–4193 (2020).
3. H. Xu and Y. Shi, "Flat-Top CWDM (De)Multiplexer Based on MZI with Bent Directional Couplers," IEEE Photonics Technol. Lett. **30**, 169–172 (2018).
4. M. He, M. Xu, Y. Ren, J. Jian, Z. Ruan, Y. Xu, S. Gao, S. Sun, X. Wen, L. Zhou, L. Liu, C. Guo, H. Chen, S. Yu, L. Liu, and X. Cai, "High-performance hybrid silicon and lithium niobate Mach–Zehnder modulators for 100 Gbit s−1 and beyond," Nat. Photonics 2019 135 **13**, 359–364 (2019).
5. A. Rao, G. Moille, X. Lu, D. Westly, M. Geiselmann, M. Zervas, K. Srinivasan, "Up to 50 dB Extinction in Broadband Single-Stage Thermo-Optic Mach-Zehnder Interferometers for Programmable Low-Loss Silicon

Nitride Photonic Circuits" 2021 Conference on Lasers and Electro-Optics (CLEO), SM1A.7, (2021).
6. I. Zand and W. Bogaerts, "Effects of coupling and phase imperfections in programmable photonic hexagonal waveguide meshes," Photon. Res. **8**, 211-218 (2020)
7. X. Qiang, X. Zhou, J. Wang, C. M. Wilkes, T. Loke, S. O'Gara, L. Kling, G. D. Marshall, R. Santagati, T. C. Ralph, J. B. Wang, J. L. O'Brien, M. G. Thompson, and J. C. F. Matthews, "Large-scale silicon quantum photonics implementing arbitrary two-qubit processing," Nat. Photonics 129 **12**, 534–539 (2018).
8. C. M. Wilkes, D. A. B. Miller, G. D. Marshall, J. Wang, J. L. O'Brien, M. G. Thompson, R. Santagati, S. Paesani, X. Qiang, and X. Zhou, "60 dB high-extinction auto-configured Mach–Zehnder interferometer," Opt. Lett. Vol. 41, Issue 22, pp. 5318-5321 **41**, 5318–5321 (2016).
9. M. Jin, J.-Y. Chen, Y. M. Sua, and Y.-P. Huang, "High-extinction electro-optic modulation on lithium niobate thin film," Opt. Lett. **44**, 1265-1268 (2019).
10. S. Liu, H. Cai, C. T. DeRose, P. Davids, A. Pomerene, A. L. Starbuck, D. C. Trotter, R. Camacho, J. Urayama, and A. Lentine, "High speed ultra-broadband amplitude modulators with ultrahigh extinction >65 dB," Opt. Express **25**, 11254-11264 (2017).
11. M. W. Puckett, K. Liu, N. Chauhan, Q. Zhao, N. Jin, H. Cheng, J. Wu, R. O. Behunin, P. T. Rakich, K. D. Nelson, and D. J. Blumenthal, "422 Million intrinsic quality factor planar integrated all-waveguide resonator with sub-MHz linewidth," Nat. Commun. **12**, 934 (2021).
12. Y.-W. Hu, Y. Zhang, P. Gatkine, J. Bland-Hawthorn, S. Veilleux, and M. Dagenais, "Characterization of Low Loss Waveguides Using Bragg Gratings," IEEE J. Sel. Top. Quantum Electron. **24**, 6101508 (2018).
13. X. Ji, S. Roberts, M. Corato-Zanarella, and M. Lipson, "Methods to achieve ultra-high quality factor silicon nitride resonators," APL Photonics **6**, 071101 (2021).
14. S. Xie, Y. Zhang, Y. Hu, S. Veilleux, and M. Dagenais, "On-Chip Fabry-Perot Bragg Grating Cavity Enhanced Four-Wave Mixing," ACS Photonics **7**, 1009–1015 (2020).
15. S. Xie, J. Zhan, Y. Hu, Y. Zhang, S. Veilleux, J. Bland-Hawthorn, and M. Dagenais, "Add–drop filter with complex waveguide Bragg grating and multimode interferometer operating on arbitrarily spaced channels," Opt. Lett. **43**, 6045-6048 (2018).
16. L. B. Soldano and E. C. M. Pennings, "Optical Multi-Mode Interference Devices Based on Self-Imaging : Principles and Applications", J. Light. Technol. **13**, 615-627 (1995)
17. A. W. Elshaari, I. E. Zadeh, K. D. Jöns, and V. Zwiller, "Thermo-Optic Characterization of Silicon Nitride Resonators for Cryogenic Photonic Circuits," IEEE Photonics J. **8**, 2701009 (2016).
18. J. F. Bauters, M. J. R. Heck, D. Dai, J. S. Barton, D. J. Blumenthal, and J. E. Bowers, "Ultralow-loss planar Si3N4 waveguide polarizers," IEEE Photonics J. **5**, 6600207 (2013).
19. D. Dai and S. Wang, "Polarization-insensitive 2x2 thermo-optic Mach-Zehnder switch on silicon," Opt. Lett. **43**, 2531–2534 (2018).
20. D. Pérez, D. Zheng, J. D. Doménech, L. Yan, W. Pan, and X. Zou, "Low-loss broadband 5x5 non-blocking Si$_3$N$_4$ optical switch matrix," Opt. Lett. **44**, 2629–2632 (2019).
21. F. Dddd, K. Cccc, D. D. Cccc, and Y. Yy, "Low-power and high-speed 2 × 2 thermo-optic MMI-MZI switch with suspended phase arms and heater-on-slab structure," Opt. Lett. **46**, 234–237 (2021).
22. T. Kita and M. Mendez-Astudillo, "Ultrafast Silicon MZI Optical Switch with Periodic Electrodes and Integrated Heat Sink," J. Light. Technol. **39**, 5054–5060 (2021).
23. R. Zhu, X. Zhou, N. Yang, L. Leng, and W. Jiang, "Towards high extinction ratio in silicon thermo-optic switches - Unravelling complexity of fabrication variation," IEEE Photonics J. **10**, (2018).
24. N. Daldosso, M. Melchiorri, F. Riboli, F. Sbrana, L. Pavesi, G. Pucker, C. Kompocholis, M. Crivellari, P. Bellutti, and A. Lui, "Fabrication and optical characterization of thin two-dimensional Si3N4 waveguides," Mater Sci Semicond Process **7**, 453–458 (2004).
25. Jiawei Wang, Kaikai Liu, Qiancheng Zhao, Andrei Isichenko, Ryan Q. Rudy, Daniel J. Blumenthal, "Fully symmetric controllable integrated three- resonator photonic molecule," arXiv: 2105.10815 (2021).